\documentclass[preprint]{iucr}              % DO NOT DELETE THIS LINE

     \journalcode{J}              % Indicate the journal to which submitted
\usepackage{amsmath}
\usepackage{amssymb}
\usepackage{color}
\usepackage[utf8]{inputenc}
\usepackage{graphicx}% Include figure files
\usepackage{dcolumn}% Align table columns on decimal point
\usepackage{csquotes}
\usepackage[T1]{fontenc}
\usepackage[caption=false]{subfig} % This was the solution for justifying the figure captions.
\usepackage{cite}
\usepackage{multicol}% http://ctan.org/pkg/multicols
\usepackage{siunitx}%typesetting units in math environments

\begin{document}                  % DO NOT DELETE THIS LINE

     %-------------------------------------------------------------------------
     % The introductory (header) part of the paper
     %-------------------------------------------------------------------------

     % The title of the paper. Use \shorttitle to indicate an abbreviated title
     % for use in running heads (you will need to uncomment it).

\title{X-ray Free Electron Laser based Dark-Field X-ray Microscopy}
%\shorttitle{Short Title}

     % Authors' names and addresses. Use \cauthor for the main (contact) author.
     % Use \author for all other authors. Use \aff for authors' affiliations.
     % Use lower-case letters in square brackets to link authors to their
     % affiliations; if there is only one affiliation address, remove the [a].

\cauthor[a]{Theodor Secanell}{Holstad}{theol@dtu.dk}{address if different from \aff}
\author[a]{Trygve Magnus}{Ræder}
\author[a]{Mads Allerup}{Carlsen}
\author[a]{Erik Bergb\"{a}ck}{Knudsen}
\author[b,c,d]{Leora}{Dresselhaus-Marais}
\author[a]{Kristoffer}{Haldrup}
\author[a]{Hugh}{Simons}
\author[a]{Martin}{Meedom Nielsen}
\author[a]{Henning}{Friis Poulsen}

\aff[a]{Department of Physics, Technical University of Denmark (DTU), Fysikvej, Building 311, 2800 Kgs. Lyngby. \country{Denmark}}
\aff[b]{Lawrence Livermore National Laboratory, Physics Division, 7000 East Ave., Livermore, CA 94550. \country{U.S.A}}
\aff[c]{Stanford University, Department of Materials Science \& Engineering, 476 Lomita Mall, Stanford, CA 94305. \country{U.S.A}}
\aff[d]{SLAC National Accelerator Lab, 2575 Sand Hill Rd, Menlo Park, CA 94025. \country{U.S.A}}

     % Use \shortauthor to indicate an abbreviated author list for use in
     % running heads (you will need to uncomment it).

%\shortauthor{Soape, Author and Doe}

     % Use \vita if required to give biographical details (for authors of
     % invited review papers only). Uncomment it.

%\vita{Author's biography}

     % Keywords (required for Journal of Synchrotron Radiation only)
     % Use the \keyword macro for each word or phrase, e.g. 
     % \keyword{X-ray diffraction}\keyword{muscle}

%\keyword{keyword}

     % PDB and NDB reference codes for structures referenced in the article and
     % deposited with the Protein Data Bank and Nucleic Acids Database (Acta
     % Crystallographica Section D). Repeat for each separate structure e.g
     % \PDBref[dethiobiotin synthetase]{1byi} \NDBref[d(G$_4$CGC$_4$)]{ad0002}

%\PDBref[optional name]{refcode}
%\NDBref[optional name]{refcode}

\maketitle                        % DO NOT DELETE THIS LINE

\iffalse
\begin{synopsis}
In this work, we demonstrate the feasibility of DFXM to image phonon wavepackets at the picosecond time scale, using an X-ray free electron laser source and a pump-probe scheme. Using the specifications of the XCS instrument at the Linac Coherent Light Source (LCLS) as an example, our simulations show the propagation of a strain wave with both a clear contrast and intensity.
\end{synopsis}
\fi

\begin{abstract}
Dark-field X-ray microscopy (DFXM) is a nondestructive full-field imaging technique providing  three dimensional mapping of microstructure and local strain fields in deeply embedded crystalline elements. This is achieved by placing an objective lens in the diffracted beam, giving a magnified projection image. So far, the method has been applied with a time resolution of milliseconds to hours. In this work, we consider the feasibility of DFXM at the picosecond time scale using an X-ray free electron laser source and a pump-probe scheme. We combine thermomechanical strain wave simulations with geometrical optics and wavefront propagation optics to simulate DFXM images of phonon dynamics in a diamond single crystal. Using the specifications of the XCS instrument at the Linac Coherent Light Source (LCLS) as an example results in simulated DFXM images clearly showing the propagation of a strain wave. 
\end{abstract}

     %-------------------------------------------------------------------------
     % The main body of the paper
     %-------------------------------------------------------------------------
     % Now enter the text of the document in multiple \section's, \subsection's
     % and \subsubsection's as required.

\section{Introduction}

During the last decade, Dark-Field X-ray Microscopy (DFXM) has emerged as a tool for mapping the microstructure within bulk crystalline materials in three dimensions \cite{Simons2016, Kutsal2019, Yildirim2020}. Using an objective lens to magnify Bragg-diffracted high-energy X-rays, DFXM facilitates mapping of orientation (with a sensitivity of 0.1 mrad) and strain (sensitivity of 10$^{-4}$ to 10$^{-5}$) in deeply embedded structures and with a spatial resolution down to 30-100 nm \cite{Poulsen2017, Poulsen2018, Kutsal2019}. This technique has been applied to characterize, e.g., grain structure and dislocation networks in metals \cite{Simons2015, Simons2016, Jakobsen2019, Dresselhaus-Marais2021}, dislocation toughening of ceramics \cite{Porz2021} and ferroelectric domains and domain walls \cite{ Simons2016, Simons2018, Schultheiss2021}. So far, DFXM has been applied at synchrotrons, with a time resolution of about 100 msec. On the one hand, this is sufficient to image some dynamic processes \textit{in situ} such as recovery in metals \cite{Yildirim2020},  dislocation motion close to the melting point in aluminium \cite{Gonzalez2020, Dresselhaus-Marais2021} and structural transformations taking place in ferroelectrics during phase transitions \cite{Ormstrup2020} or mechanical loading. While DFXM was recently demonstrated at an X-ray free electron laser (XFEL), its time resolution measuring X-ray damage was still limited to 33 ms by the repetition rate of the source \cite{Dresselhaus-Cooper2020}. On the other hand, other dynamic processes, such as phonon propagation, occur on much shorter picosecond timescales. Up to now, imaging of phonon waves has been limited to thin specimens ($\lesssim$ 50 nm) using dark-field transmission electron microscopy \cite{Cremons2016}, to nanocrystals using Bragg coherent diffraction imaging \cite{Clark2013, Wen2019}, and to optical imaging \cite{Ofori-Okai2014}. 

An optical-laser-pump / X-ray-probe scheme has previously been applied to study phonon dynamics using diffraction methods at XFELs, where the enhanced brightness and pulse duration allow for finer time-resolutions necessary to study phonon dynamics \cite{Lindenberg2000, Larsson2002,  Persson2015, Jarnac2017, Lemke2018}. Here, we propose to implement DFXM at an XFEL using such a pump-probe scheme. This will allow direct visualization of the interaction between sound waves and microstructural features such as domain walls or dislocations deep within mm-sized single- and poly-crystalline materials - of interest to materials physics, seismology and extreme condensed matter science.  

In this work, we provide a numerical demonstration of the feasibility of this scheme. We simulate longitudinal sound wave propagation in a diamond single crystal using the  thermomechanical Python modeling package \texttt{udkm1Dsim} \cite{Schick2014, Schick2021}. Forward simulations of DFXM data are performed using both geometrical optics \cite{Poulsen2021} and wavefront propagation optics \cite{Carlsen2021a} to explore the DFXM parameter-space and optimize the experimental configuration. Using a setup relevant for an experiment at the XCS instrument at the Linac Coherent Light Source (LCLS) as an example, the simulations show the propagation of a longitudinal strain wave with both a clear contrast and a good signal-to-noise from a single pulse, thereby opening the door to investigating the materials science entailed in dynamic imaging of phonon propagation \cite{Wolfe1995, Siemens2010, Hatanaka2014}. Moreover, we demonstrate how geometrical aspects of this experiment may be optimized to improve contrast.

\section{Experimental }
\subsection{DFXM geometry and contrast}

The methodology of DFXM in general and associated properties such as spatial and reciprocal space resolution are presented in detail in e.g. Refs.~\cite{Poulsen2017, Poulsen2018, Poulsen2021}.

\begin{figure}
\includegraphics[width=0.95\textwidth]{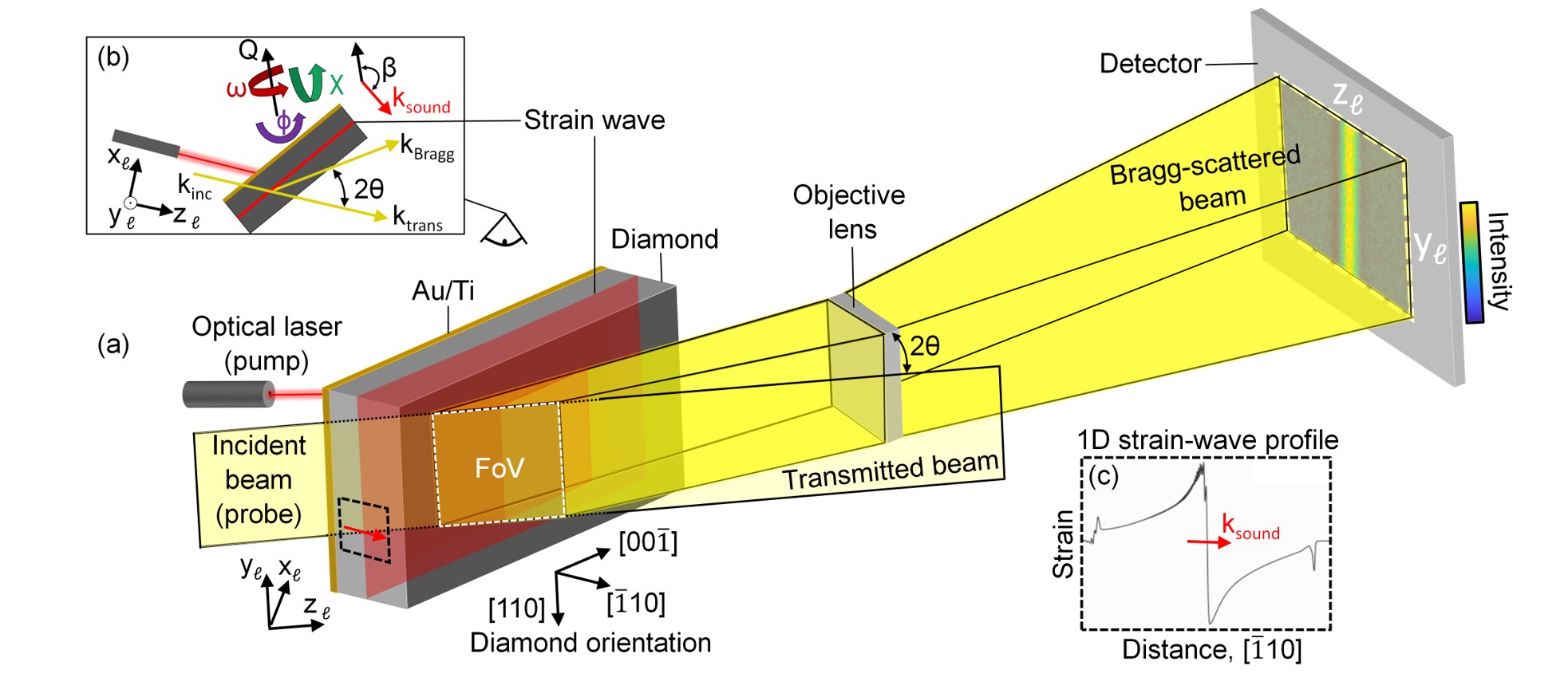}
\caption{(a) Schematic of the configuration for pump-probe DFXM imaging of a strain wave used for the simulations. A crystallographic coordinate system indicates the diamond orientation. An optical laser (pump) heats a deposited Au film. This leads to thermal expansion and the launching of a strain wave (red plane) into diamond. An incident X-ray beam (probe) is condensed into a sheet which penetrates the diamond crystal at some time delay after the laser excitation. A laboratory coordinate system $(x_\ell,y_\ell,z_\ell)$ is defined with $x_\ell$, $y_\ell$ and $z_\ell$ parallel to the incident beam width, height and propagation direction, respectively. The diamond crystal is oriented such that the \{111\}-planes Bragg scatter, and an objective lens is positioned in the Bragg scattered beam. This results in a magnified DF image of a field of view (FoV) (white dashed lines) being projected onto the detector. The colorscale shows the photon count in a simulated DF image (see section \ref{subsec:DFXM_modelling}). (b) The upper left inset shows a top-down view of the setup. We assume the sample is mounted on a goniometer which can perform  
an $\omega$-rotation around the local diffraction vector $Q$ and two orthogonal tilts, $\chi$ and $\phi$. The angle $\beta$ between the wave-vector of the strain wave $k_{\textrm{sound}}$ and the diffraction vector $Q$ is indicated. (c) A profile of the simulated 1D strain-wave profile (see section \ref{subsec:sw_modelling}) is shown in the inset with dark dashed lines.}
\label{fig:fig1}
\end{figure}

The DFXM geometry considered in this work is illustrated in Fig.~\ref{fig:fig1}\footnote{Note that in the XFEL community, the optical axis is along $z_\ell$, whereas in the DFXM community it typically is along $x_\ell$.}. A nearly monochromatic  X-ray beam with wavelength $\lambda$ is condensed in the horizontal direction to  generate a vertical line beam focused onto the sample; this defines a vertical plane of observation inside the sample. The incident beam is defined by its vertical width (i.e. height) $\Delta y$, its horizontal thickness $\Delta x$, and its horizontal divergence $\Delta \zeta_h$. 

The sample is mounted on a goniometer designed to access diffraction angles in a  horizontal scattering geometry, and probe reciprocal space  in the immediate vicinity of a given reflection $\vec{Q}_0$, corresponding to lattice planes $(h,k,\ell)$. The implementation considered here achieves this by moving the sample along a combination of $\omega, \chi$ and $\phi$ rotation stages, see Fig.~\ref{fig:fig1} (b). The direction of the diffracted beam in the horizontal plane is characterized by the scattering angle, $2\theta_0$ (for the nominal $(h,k,\ell)$ reflection). The motor-position in the horizontal plane is denoted as $2\theta$ (Fig.~\ref{fig:fig1}). We shall assume that the motor-position is exactly at the scattering angle, i.e. that $2\theta = 2\theta_0 $. %\footnote{For ease of representation, in the following the definition of $\eta = 0$ is different from that of Refs. \cite{Poulsen2017, Poulsen2018, Poulsen2021}, where a vertical scattering geometry is used.}. 
The optical axis of an X-ray objective lies along the diffracted beam for  $\vec{Q}_0$  to produce a magnified and inverted image of the illuminated plane on the 2D detector. As is usually used for DFXM experiments, we assume a Compound Refractive Lens (CRL) objective for this simulation \cite{Snigirev1996}, which is a thick lens comprising $N$ identical parabolic lenslets \cite{Simons2017}. The CRL is characterised by a Numerical Aperture, $NA$, and a focal length $f_N$.  The image generated by the objective has an associated magnification, $\mathcal{M}$, and field of view (FoV) in the object plane (i.e. the crystalline sample). The projection angle implies the illuminated plane is observed at an angle with a resulting aspect ratio of 1:$\tan(2\theta)$. 

The relation between reciprocal space and  strain components and between micro-mechanical models and DFXM images are discussed in detail in Ref.~\cite{Poulsen2021}. The vicinity of $\vec{Q}_0$ in reciprocal space can be probed in three orthogonal directions by scanning $\phi$, $\chi$ and a combination of $\phi $ and $2\theta$, respectively. These are referred to as rocking, rolling (both of which probe the local mosaicity) and longitudinal strain scans, respectively, and may be combined.  The contrast in the resulting images scale based on the local values of three of the displacement gradient tensor field elements. The longitudinal $\phi - 2\theta$ scan  corresponds to probing the axial strain along the diffraction vector $\vec{Q}_0$. For a material with no rotation of the lattice planes the two others represent shear strains. 

\subsection{Dynamic strain wave characterization}
 
The aim of this article is to provide insight into how to design and optimize an optical-laser-pump - X-ray-probe DFXM experiment at an XFEL with the aim of imaging strain wave movement. At XFELs, a single pulse typically has a duration of 1-100 fs \cite{Inoue2019}. Provided  the contrast and signal-to-noise is suitable, one may then image the structure and strain in the illuminated layer, averaged over this time interval. 
Here, we focus on reversible processes that are suitable for a pump-probe scheme. Specifically, we consider the visualisation of an isolated longitudinal sound wave traversing a single crystal. With sound velocities of 18 km/s \cite{Wang2004}, the strain wave will travel 18 pm - 1.8 nm  within the duration of the pulse, much less than the spatial resolution, meaning that no temporal blurring needed to be considered in the image.

The longitudinal sound wave will give rise to one strain component $\varepsilon$ in the direction of its propagation direction, $\vec{k}_{\textrm{sound}}$. Let $\beta$ be the angle between $\vec{Q}_0$ and $\vec{k}_{\textrm{sound}}$ (Fig.~\ref{fig:fig1}). Then, the projection $\varepsilon \cos(\beta)$ is probed in the DFXM experiment. Depending on the goniometer setting, the projected strain can be visualized in several ways, including ``strong-beam'' and ``weak-beam'' conditions \cite{Jakobsen2019} to be explored below.

\subsection{Experimental details used in the simulations}
\label{sub-expdetails}

To determine the feasibility of visualising sound waves  we shall consider propagation of such a wave in single crystal diamond with an experimental set up that is relevant to the XCS instrument at the LCLS. That instrument is well suited for DFXM as it is equipped with a long arm (8 m) that is rotated radially about a sample's diffraction $2\theta$ angles. The layout of such an experiment is shown in Fig.~\ref{fig:fig1}. To ensure that the results of the numerical simulations will reflect future experimental realities, we describe the detailed parameters that are considered typical of this beamline. 

The diamond single crystal is specified to have dimensions of 0.6 mm $\times$ 1 mm $\times$ 2 mm with ($1\bar{1}0$), (110) and (001) facets. The ($1\bar{1}0$) facet is coated with a 15 nm Ti adhesion layer followed by a 600 nm Au layer, which thermalizes when irradiated optically. We shall assume a width of the diamond sample rocking curve of $ \Delta_{\textrm{mosaic}} = 200\,\si{\micro rad}$. The Au layer is  excited with a 100\,fs laser pulse ($\lambda$ = 800 nm; fluence = 0.8 J/cm$^2$). The subsequent impulsive expansion of the Au film results in strain waves propagating from the Au film and into the diamond crystal (see section \ref{subsec:sw_modelling} below).

The  XFEL pulses are assumed to be 10 keV, with a pulse duration of 35 fs and energy of  2 mJ. These are monochromatized to give an energy band width (FWHM) of $\Delta E /E$ $\sim$ $10^{-4}$ and a divergence of approximately 1.1 x 1.1 µrad$^2$ after the monochromator.  Before the beam hits the sample, it is condensed by a set of 1D Be-lenses into a line beam (Fig. \ref{fig:fig1}) with vertical width $\Delta y$ = 500 µm,  a horizontal thickness (FWHM)  of $\Delta x  = 3$ µm, and a horizontal divergence (FWHM) of $\Delta \zeta_h$ = 30 µrad. The $10^{-4}$ bandwidth of the monochromator and the significant width, intra pulse substructure due to microbunching and shot-to-shot variation in $I(E)$ in combination with absorption loses in the lens system means that the average photon count after the monochromator can be expected to be reduced to  $N_{\textrm{estimated, inc}} = 2 \cdot 10^{10}$ photons/pulse incident on the sample. As such, this is the average photon count used in the simulations below, but we note that significant shot-to-shot variation in photon count (from 0 to 10$^{12}$ ph/pulse) is expected in an actual XFEL experiment.

The incoming X-ray sheet slices through the crystal, and the X-rays that are Bragg-scattered from \{111\}-planes (2$\theta_0$ = 35.04$^{\circ}$) are  projected by the objective lens stack onto a 2D detector at a fixed position $L = 7.1$ m downstream from the sample. The objective is a CRL with $N = 30$ Be lenslets, each with a radius of curvature of $R =$ 50 µm, a distance between centers of neighboring lenslets of $T=2$ mm, and a thickness of each lenslet of $T_{\textrm{effective}} = $ 1 mm. Using the analytical expressions from Ref. \cite{Simons2017} the corresponding effective focal length is $f_N = 0.207$ m, the sample-to-objective-entry plane distance $d_1 = 0.215$ m, objective-to-detector distance $d_2 = 6.83$ m, the effective numerical aperture (FWHM) $NA= 0.000845 $, the physical aperture 447 µm and the magnification of the X-ray objective $\mathcal{M} = 27.9 $. The CMOS detector at the beamline is assumed to have a pixel size of 6.5 x 6.5 µm$^2$ and 2560 x 2160 pixels. In the simulation below we assume a 1:1 coupling between the scintillator screen and the camera, and 2x binning. The effective pixel size in the observation plane (the intersection between the incident beam and the crystal) is then 466 x 664 nm$^2$.

\section{Modelling}

The full DFXM phonon modelling begins with the sample's thermomechanical model of the light-matter interaction and associated phonon wavepacket, i.e. the strain wave. The resulting micro-mechanical model is then used as the input for forward models of the DFXM images as function of $(\phi, \chi, 2\theta)$  and of the  orientation of the single crystal.      
The parameters of the experimental configuration  presented in Section
\ref{sub-expdetails} are used throughout.

\subsection{Strain wave modelling}
\label{subsec:sw_modelling}

A laser is used to excite an Au film deposited on a diamond crystal (see section \ref{sub-expdetails} for details). A one-dimensional thermo-mechanical model, \texttt{udkm1Dsim} \cite{Schick2014, Schick2021}, is utilized to compute the propagation of the resulting strain wave in the propagation direction, $z_{\textrm{sw}}$. This model has previously been successfully implemented to simulate strain wave propagation in a gold-coated indium antimonide crystal \cite{Jarnac2017}.

The \texttt{udkm1Dsim} Python package uses  the 2-temperature model \cite{Jiang2005, Tzou2002}, where the electrons and lattice have separate temperatures and heat diffusion equations, to compute the temperature distribution after the laser-excitation. For the electrons, the heat diffusion equation is 

\begin{equation}
\label{eq:1}
C_e(T_e)  \frac{\partial T_e}{\partial t} =  \frac{\partial}{\partial z_{\textrm{sw}}} \left(k_e(T_e, T_l) \frac{\partial T_e}{\partial z_{\textrm{sw}}} \right) + G_e(T_e, T_l) + S(z_{\textrm{sw}},t).
\end{equation}

Here, $T_e$ is the electronic temperature,  $C_e$ is the heat capacity of the electrons, and $k_e$ is the thermal conductivity of the electrons. $S(z_{\textrm{sw}},t)$ is the source term which describes absorption of the laser energy by the electrons (see Ref.~\cite{Schick2014} for details), and $G_e$ is the lattice-electron coupling factor.

The corresponding heat diffusion equation for the lattice ($l$) is

\begin{equation}
\label{eq:2}
C_l(T_l)  \frac{\partial T_l}{\partial t} =  \frac{\partial}{\partial z_{\textrm{sw}}} \left(k_l(T_l) \frac{\partial Tl}{\partial z_{\textrm{sw}}} \right) + G_l(T_e, T_l). 
\end{equation}

 The lattice heat capacity and thermal conductivity  are  temperature independent, while the temperature dependence of the electronic heat capacity and thermal conductivity are given by \cite{Tzou2002}
\begin{equation}
\label{eq:3}
C_e(T_e)  = C_{e0} \left( \frac{T_e}{T_0} \right),
\end{equation}
and
\begin{equation}
\label{eq:4}
k_e(T_e, T_l)  = k_{e0} \left( \frac{T_e}{T_l} \right),
\end{equation}
respectively. Here $T_0$ is the initial temperature (300 K). 

The electron-lattice coupling terms in the diffusion equations are taken to be proportional to the difference between the electronic and lattice temperatures \cite{Tzou2002, Jiang2005, Schick2014, Jarnac2017, Schick2021}: 

\begin{equation}
G_l(T_e, T_l) = G(T_e - T_l)
\end{equation}

\begin{equation}
G_e(T_e, T_l) = G(T_l - T_e)
\end{equation}

The proportionality factor $G$ is also temperature dependent \cite{Jiang2005, Lin2008}, and the temperature dependencies used in the present simulations are taken from Ref.~\cite{Lin2008}.

Solving the heat diffusion equations (\ref{eq:1},\ref{eq:2}) gives a temperature profile at different time delays after the laser excitation.  The linear thermal expansion coefficient of the lattice ($\alpha$ in table \ref{table:sw_params}) is then used to compute the thermal expansion. Subsequently, a model comprised of a linear chain of point masses connected by springs is used to compute the resulting lattice dynamics of the longitudinal strain waves - see Ref.~\cite{Schick2014} for details.

\begin{table}
\caption{Parameters used in strain-wave simulations.}
\label{table:sw_params}
\tiny
\begin{tabular}{llll}
 Properties  & Gold  & Ti & Diamond  \\ \hline 
 $k_{e0}$ [Wm$^{-1}$K$^{-1}$] & 315$^{\dagger}$    & 10$^\parallel$ & 0 \\
 $k_{l}$ [Wm$^{-1}$K$^{-1}$] & 2.6$^{\ddagger}$ & 10$^\parallel$ & 1200* \\
 $C_{e0}$ [Jm$^{-3}$K$^{-1}$] & 2.1 $\cdot$ 10$^4$$^{\dagger}$ & 328.9$^\mathsection$ & 0 \\
 $C_{l}$ [Jm$^{-3}$K$^{-1}$] & 2.5 $\cdot$ 10$^6$$^{\dagger}$ & 2.21 $\cdot$ 10$^6$$^{\dagger \dagger}$ & $^\mathparagraph$ \\ 
 $G$ [Wm$^{-3}$K$^{-1}$] & $^\mathsection$ & $^\mathsection$ & $^\mathsection$ \\
 $c_s$ [nm/ps] & 3.24$^{\ddagger \ddagger}$ & 6.16** & 18$^{\mathsection \mathsection}$ \\
 $\alpha$ [K$^{-1}$] $\cdot$ 10$^{6}$ & 13.99$^{\mathparagraph \mathparagraph}$  & 8.5***  & 1.06$^{\mathsection \mathsection \mathsection}$ \\
 $\rho$ [g/cm$^3$] & 19.3 & 4.506 & 3.51$^{\dagger \dagger \dagger}$ \\

 $\dagger$\cite{Tzou2002}, $\parallel$\cite{Klemens1986}, $\ddagger$\cite{Wang2016}, & & \\
 *\cite{Anthony1990}, $\mathsection$\cite{Lin2008}, $\mathparagraph$\cite{Victor1962}, & & \\
 $\dagger \dagger$\cite{Rao1979}, $^{\ddagger \ddagger}$\cite{Ismail2018}, **\cite{Zaretsky2008}, & & \\
 $\mathsection \mathsection$\cite{Wang2004}, $\mathparagraph \mathparagraph$\cite{Dutta1963}, ***\cite{Hidnert1943}, & & \\ 
$\mathsection \mathsection \mathsection$\cite{Jacobson2019}, $\dagger \dagger \dagger$\cite{Graebner1996} 
\end{tabular}
\end{table}

\subsection{DFXM forward modelling}
\label{subsec:DFXM_modelling}

In the following, we consider DFXM imaging based on the most intense reflection $\vec{Q}_{0} =<$111$>$. Moreover, we will assume that the propagation direction of the sound wave, $\vec{k}_{\textrm{sound}}$, is in the scattering plane at an angle $\beta $ to $\vec{Q}_{0}$ (Fig.~\ref{fig:fig1}).

\subsubsection{Geometrical optics formalism}

A formalism for forward simulation of DFXM images based on geometrical optics is described in Ref.~\cite{Poulsen2021}. The input is a voxelized version of a displacement gradient tensor field - which in our case is readily provided by the 1D strain wave model above. The forward simulation is based on analytical expressions for the relation between strain, reciprocal space and the resulting detector intensity distribution. The instrumental resolution function is determined by Monte Carlo simulations prior to the actual forward simulations. This means that the method is relatively fast and well-suited for optimization purposes.  On the other hand, geometrical optics is not inherently suited for studying effects such as coherence  and in particular dynamical diffraction, which may be relevant for DFXM studies of single crystals.     

The simulations below were performed by a slightly revised version of the code provided as Supplementary Information to Ref. \cite{Poulsen2021} that also now takes into account counting noise. This is based on a model of the counts per pixel for the  strain-free single crystal in the optimised ``strong-beam'' condition, the latter term meaning that the crystal’s orientation with respect to the incident  beam  is  centered on the  Bragg  condition,  i.e. the strongest part of the rocking-curve ($\phi = \chi = 0, 2\theta = 2\theta_0$). Specifically, the diffracted photon count for one pulse as summed over the image for the strong-beam setting is estimated as follows

\begin{eqnarray}
N_{\textrm{estimated, strong}} = N_{\textrm{estimated, inc}} \frac{A_{\textrm{FoV}}}{A_{\textrm{ill}}} T \frac{\Delta_{\textrm{Darwin}}}{\Delta_{\textrm{mosaic}}} \frac{1}{2}.
\label{eq-intensity-cal}
\end{eqnarray}

Here, $N_{\textrm{estimated, inc}}$ is the estimated total number of photons incident on the sample, $A_{\textrm{FoV}}$ is the area of the field-of-view in the illuminated plane, and $A_{\textrm{{ill}}}$ is the total area of the illuminated plane within the diamond crystal: $A_{\textrm{ill}} = \Delta y \cdot b/\sin[\angle(k_{\textrm{inc}},\textrm{crystal surface})]$, where $[\angle(k_{\textrm{inc}},\textrm{crystal surface})]$ is the grazing angle between the incident X-ray beam and the diamond crystal's surface and $b$ is the diamond crystal's thickness along the direction perpendicular to the gold coated facet. $T$ is the average transmission through Au and diamond, and $\Delta_{\textrm{Darwin}}$ is the Darwin width, which for diamond(111) and 10 keV is 20 µrad.  $\frac{\Delta_{\textrm{Darwin}}}{\Delta_{\textrm{mosaic}}}$ is a rough approximation for the ratio of the sample diffracting dynamically and the factor $\frac{1}{2}$ expresses that in this regime, the average photon count is equally distributed between the transmitted and Bragg-scattered beams. We will in the simulations below work with a region-of-interest of $M$ x $M$ = 100 x 100 binned pixels.

Next, a simulated value of the strong-beam photon count is calculated for the strain-free single crystal, $N_{\textrm{simulated, strong}}$. The number of photons detected in each pixel is then determined by sampling from a Poisson distribution with mean given by the pixel values in raw simulated images, $N_{\textrm{simulated, raw}}(i,j)$ (as in Ref.~\cite{Poulsen2021}) scaled by the ratio of the estimated to the simulated strong-beam value for the diffracted photon count,
\begin{equation}
\label{eq:noise}
    N_{\textrm{abs}}(i,j) \in \mathrm{\textrm{Pois}}\left(\frac{N_{\textrm{estimated, strong}}}{N_{\textrm{simulated, strong}}}\cdot N_{\textrm{simulated, raw}}(i,j)\right).
\end{equation}
The scaling factor in equation (\ref{eq:noise}) relates the raw photon count of the simulated image $ N_{\textrm{simulated, raw}}$ to the photon count estimated in equation (\ref{eq-intensity-cal}). This scaling is valid for each pixel, because in a strain-free single crystal the image is homogeneous and the diffracted photon count per pixel is $N_{\textrm{{\textrm{total}}}}/M^2$, where $M^2$ is the total number of pixels. This scaling is also used for the so-called ``weak-beam'' conditions - where the crystal's orientation with respect to the incident beam is just at the edge of the Bragg condition, i.e. the weakest part of the rocking-curve - because the simulations internally scale weak and strong-beam conditions correctly. For clarity, we note that this procedure does not take any non-ideal effects of the detector into account. It merely yields a realistic estimate of the signal arriving at the detector plane. The noise model is included as Supplementary Information.

\subsubsection{Wave  optics formalism}
% advantage of wave-optics
The wave-optics simulations represent linearly polarized X-rays as a coherent wave front of an electric field with the complex-valued amplitude given on a discrete two-dimensional grid.
The electric field throughout the crystal is simulted as two components, one propagating along the incident beam and one propagating in the scattered beam direction.
Scattering is handled by the transfer of amplitude between the incident and scattered electric field, with scaling based on the electron density of the deformed crystal lattice in each voxel ($\approx$structure factor of $(h,k,\ell)$ and $(\bar{h},\bar{k},\bar{\ell})$) in the formalism of the Takagi-Taupin equations defined in Ref.~\cite{Takagi1962}.
The Takagi-Taupin formalism allows for multiple scattering events, i.e. dynamical diffraction, where each scattering event introduces a 90$^\circ$ phase shift.
The images in the strong-beam condition are dominated by the so-called Pendellösung fringes that arise from multiple scattering events, as intensity beats between the direct and the scattered beam (i.e. interference effects from the phase shift introduced by the scattering). 
The wave-optics formalism therefore allows us to investigate the strong-beam condition, as well as verifying that offsetting the sample by rotating the goniometer in $\phi$ yields only weak-beam contrast, with negligible contribution from dynamical diffraction.

% description of the implementation for the scattering
The Takagi-Taupin equations are numerically integrated using the method given in Ref.\cite{Carlsen2021a}. % description of the implementation for the CRL 
The scattered field is propagated  through the CRL and to the detector using Fourier propagation methods described in Ref.~\cite{Pedersen2018}. A 3D grid of $16000\times1200\times8000$ voxels is used, with a voxel size of 125$\times$62.5$\times$62.5 nm$^3$ along the $[00\bar{1}]$, [110] and [$\bar{1}10$] directions, respectively. The full size of the diamond crystal along $[00\bar{1}]$ and $[\bar{1}10]$ is simulated in order to correctly capture the fringes caused by dynamical diffraction, but we do not simulate the full size of the crystal along [110], as the crystal is taller than the FoV. A Gauss-Schell beam is used, which is a quasi-monochromatic beam that has a Gaussian profile in both real and reciprocal space represented by a series of (independently coherent) modes. The majority of the intensity is contained in the first mode (a largely coherent beam), and only 4 modes are included. The parameters of the Gauss-Schell beam are $\sigma_I = 1.7$~µm and $\sigma_c = 1.2$~µm as defined in Ref.~\cite{Starikov1982}.

\section{Results} 

\subsection{Strain wave modelling}

%The results of our simulations of strain waves in the coated diamond,  using the parameters presented in section \ref{sub-expdetails} and listed in table \ref{table:sw_params}, is shown in Fig.~\ref{fig:fig2}.

\begin{figure}%[h!]
    \centering
    \includegraphics[scale=0.65]{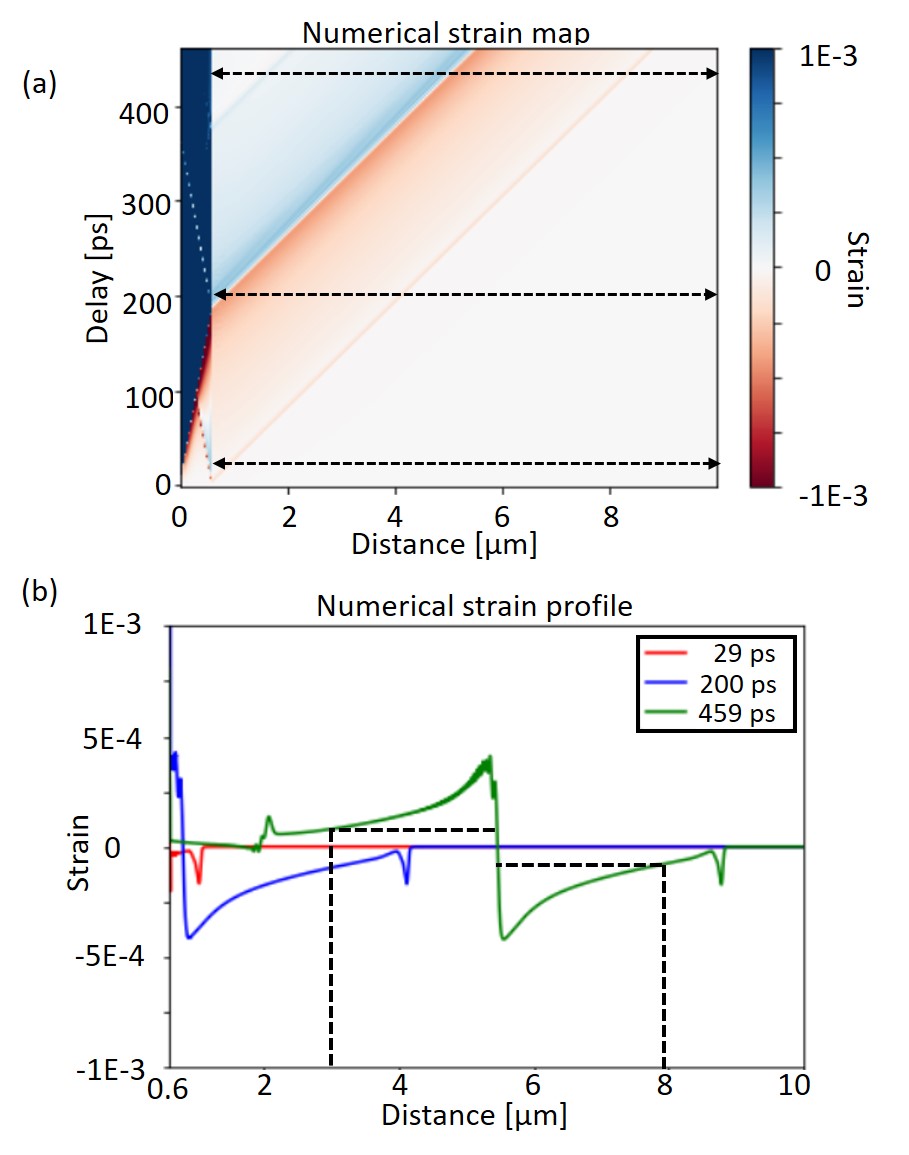}
    \caption{Strain wave profile in a diamond single crystal as function of time delay from laser pulse heating, as computed using a 1D thermomechanical model. (a) 2D map of strain versus depth and time delay. (b) 1D plots of the strain profile in diamond at different time delays (indicated with dashed lines in (a)). The spatial extent of the part of the strain wave that is visible in DFXM at 459 ps (indicated with dashed lines) is about 5 µm.}
    \label{fig:fig2}
\end{figure}

Figure \ref{fig:fig2} (a) displays the strain profile with distance and time on the horizontal and vertical axis, respectively. Figure \ref{fig:fig2} (b) shows the strain inside of diamond as a function of distance from the gold film, with different colors showing different time delays. 

The traces furthest from the surface (i.e. later in time delay) demonstrate that the strain profile is anti-symmetric around the "position" of the strain pulse with a maximum strain that is approximately 4 $\times$ 10$^{-4}$. Given the strain resolution of DFXM, the spatial extent of the visible part of the strain wave is about 5 µm (see Fig.~\ref{fig:fig2} (b)). In diamond the wave travels 6.3 Å during the LCLS XFEL pulse duration (35 fs), which is much smaller than the spatial resolution and thus negligible. %The strain profile input to the X-ray forward simulations was a slightly simplified version of this.

\subsection{DFXM forward modelling}
\subsubsection{Geometrical optics formalism}
\begin{figure}
\centering
\includegraphics[width = 0.78\textwidth]{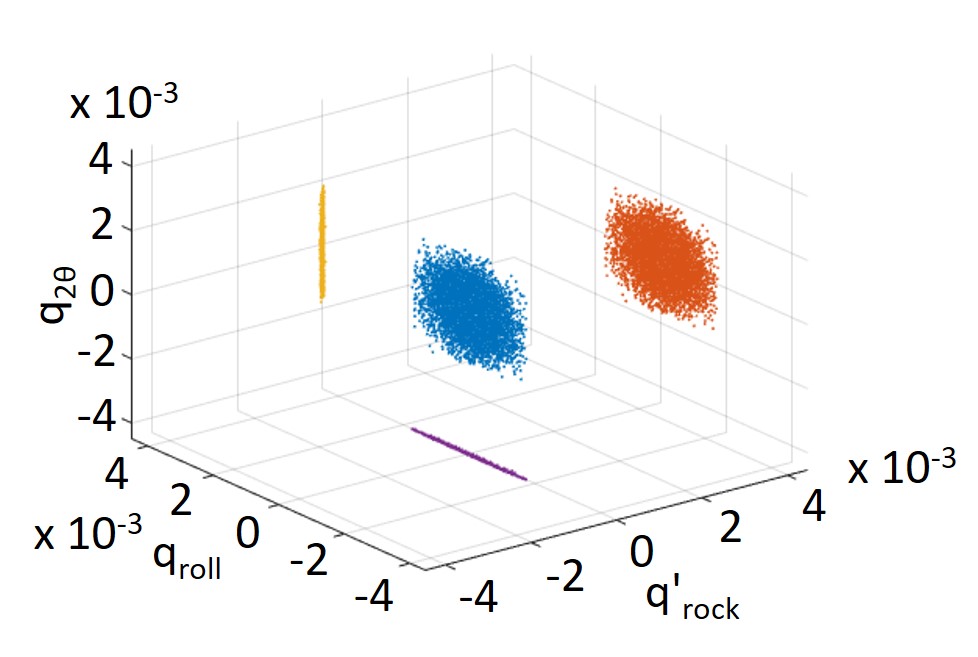}
\caption{Reciprocal-space resolution function for the simulated DFXM setup. The simulation involved   10000 simulated rays. Blue symbols: 3D scatter plot for the resolution function expressed in a coordinate system colinear with the image plane. The purple, orange and yellow symbols correspond to 2D projections onto the $q_{\mathrm{rock'}}$-$q_{\mathrm{roll}}$ plane, $q_{\mathrm{roll}}$-$q_{2\theta}$ plane and the $q_{\mathrm{rock'}}$-$q_{2\theta}$ plane, respectively. }
\label{fig-recspace}
\end{figure}

From the experimental geometry introduced above, the resolution as expressed in terms of how a single point in $q$-space is broadened due to finite $NA$s etc. can be estimated. This broadening is referred to as the \emph{reciprocal space resolution function} and Fig.~\ref{fig-recspace} shows this based on Monte Carlo modelling of 10000 simulated rays propagating through the optical system for the  $\phi = \chi = 0$ and $2\theta = 2\theta_0$ scattering geometry\footnote{Strictly speaking, the simulation is only valid for a sample position which is on the optical axis of the objective. See Ref.~\cite{Poulsen2021}, which also presents expressions for the general case.}. The Monte Carlo ray simulation is visualised in a $(q_{\mathrm{rock'}}, q_{\mathrm{roll}}, q_{2\theta})$ coordinate system, which is referred to as the ``imaging coordinate system,'' in Ref. \cite{Poulsen2021}.  It is rotated with respect to the laboratory system shown in Fig.~\Ref{fig:fig1} by $2\theta$ around the $y_\ell$-axis. 

The 3D resolution function (blue) in Fig.~\ref{fig-recspace} is projected onto the three possible $q$-space planes to demonstrate the intrinsic anisotropy, which is significantly more dramatic than is observed for synchrotron experiments \cite{Poulsen2021}. Comparison of the projection shown in orange with the ones in yellow and purple shows a large anisotropy in the reciprocal-space resolution function. To first order, the resolution function is a disc, with a ``thin dimension'' parallel to the optical axis of the objective. The width (FWHM) of this is below $10^{-4}$. The dimensions of the two wide axes are defined by the acceptance functions set by the $NA$ of the objective lens, producing a nearly planar distribution. The width in the $q_{\textrm{rock'}}$ direction is largely determined by the horizontal divergence of the incident radiation, i.e. the $NA$ of the condenser lens, and is not the result of a more coherent source.

The majority of the crystal is strain free and will therefore give rise to diffraction at the setting where $\vec{Q} = \vec{Q}_0$. The sound wave may be visible in this strong-beam condition, but dynamical diffraction makes it difficult to quantify and interpret such images, and in any case such contrast cannot readily be simulated by geometrical optics.

By rotating the sample in $\phi$  -- rocking the sample -- by an amount $\Delta \phi$, which is larger than the thin dimension of the resolution function (moving along $q_{\textrm{rock'}}$ in Fig.~\ref{fig-recspace}), dynamical scattering from the unstrained part of the crystal is avoided because it no longer satisfies the Bragg condition. Such a movement corresponds to a shear strain of magnitude $ \Delta \phi$. Likewise, by rotating the sample  in $\chi$  - rolling the sample (moving along $q_{\textrm{roll}}$ in Fig.~\ref{fig-recspace}) – by an amount $\Delta \chi$, which is larger than $NA/(2\sin(\theta))$, dynamical scattering from the unstrained part of the crystal is avoided, and we have another type of weak-beam contrast. Such a movement corresponds to a shear strain of magnitude $\Delta \chi$ \cite{Poulsen2017}.

Finally, by offsetting the crystal in $2\theta$ by an amount $\Delta 2\theta$ larger than $NA$ - and a simultaneous offset in $\phi$ - weak-beam contrast is obtained in the longitudinal direction (moving along $q_{2\theta}$ in Fig.~\ref{fig-recspace}). In this case the movement corresponds to an axial strain of magnitude $\Delta 2\theta/ (2\tan(\theta))$ \cite{Poulsen2017}.

\iffalse
\color{red}[You explain a lot of different versions of weak-beam here and say "this is weak-beam", but it is really hard to follow (a) if those discussions are general or are describing a figure/simulation you are showing us, and (b) how this story connects to your description of weak-beam contrast earlier in this manuscript, which you describe based on a CRL alignment offset. I think these paragraphs need to be re-worked. Also, you should go through the full paper. You introduce weak/strong-beam contrast many times, and do it differently each time. This makes it hard to follow how the new information should change our understanding of the previous description. -L]\color{black}

To describe the contrast quantitatively we introduce a reciprocal-space resolution function, based on the Monte-Carlo methods described above(?): for a given instrumental setting $(\phi_0,\chi_0, 2\theta_0)$ this is defined as the normalised intensity on the detector as function of scattering in direction $(\phi,\chi, 2\theta)$. A simulation of the reciprocal space function is displayed in Fig.  \ref{fig-recspace}. It appears that it is anisotropic, implying that the lower limits on strain that can provide contrast varies substantially. \color{red}[This whole paragraph seems to restate what is above, but in a different way that are making me unclear if I understand what you are doing in Fig. 3. -L]\color{black}
\fi

From the resolution function in Fig.~\ref{fig-recspace}, we anticipate to resolve the sound wave in more detail when weak-beam contrast can be obtained by rocking the sample (because the resolution function is narrow along $q_{\textrm{rock'}}$). For rolling and longitudinal strain scans we note that the diameter of the disc in reciprocal space nearly matches the full range of the strain field. For this reason, it is difficult to reach weak-beam contrast for such scans. Instead it may be relevant to operate in a strong-beam contrast mode, where the contrast is reversed. 

\begin{figure}
\centering
\includegraphics[width = 0.99\textwidth]{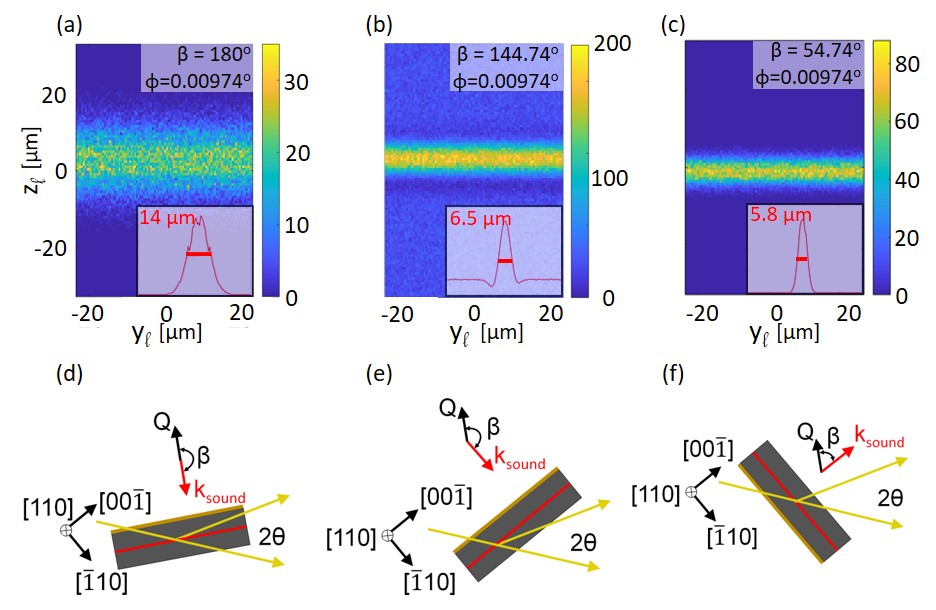}
\caption{Single pulse DFXM images simulated using geometrical optics at the time when the sound wave is centered within the FoV. The three rows plot cases in which the experiment is set up with angles between diffraction vector and sound propagation direction of (a) $\beta = 180^{\circ}$, (b) $144.74^{\circ}$, and (c) $54.74^{\circ}$, respectively (corresponding to sound waves propagating along [$\bar{1}$11] and [$\bar{1}$10], and [00$\bar{1}$], respectively), as indicated schematically in the bottom row (d-f). All DFXM images have an orientational offset in rocking direction $\phi$ to achieve weak-beam contrast. The pixel size in the object plane (intersection between incident beam and crystal, i.e. the \textit{gauge volume}) is 466 x 664 nm$^2$. The colorbar denotes number of detected photons per pixel, without corrections for quantum efficiency or other detector effects. The insets are integrated photon-count profiles along the $z_{\ell}$-direction, and the FWHM is indicated in red.}
\label{fig:fig4}
\end{figure}

We have forward simulated DFXM images of the strain wave propagation for a number of angles $\beta$. In Fig. \ref{fig:fig4}, we present snapshots for $\beta = 180^{\circ}$, $144.74^{\circ}$, $54.74^{\circ}$ corresponding to the strain wave propagating along [$\bar{1}$11] and [$\bar{1}$10], and [00$\bar{1}$], respectively. Note that Fig. \ref{fig:fig4} (a) corresponds to a diamond with different facets and Fig. \ref{fig:fig4} (c) a diamond with different shape compared to the one shown in Fig.~\ref{fig:fig1}. We demonstrate the use of two types of contrast: rocking-type weak-beam contrast and rolling-type strong-beam contrast. Suitable offsets were identified by inspection to be $\Delta \phi =  0.00974^{\circ}$ and $\Delta \chi = 0.0974^{\circ} $, respectively. Corresponding movies of the entire $\phi$ and $\chi$ scans are available as Supplementary material.

The $\phi$-scans show strong contrast in the weak-beam condition, with pixel intensity differentials across the strain-wave as high as 200 counts. For all $\beta $ cases, weak-beam contrast appears  in the images when the offset in $ \phi$ is in the range between the max strain, $4 \times 10^{-4}$ rad, and the width of the reciprocal-space resolution function in direction $\phi$, c.f. Fig.~\ref{fig-recspace}. 

The variations with $\beta$ are readily explained. For $\beta = 180^{\circ}$, the planar wave front is more parallel to the observation plane in the crystal, leading to a wider and consequently less intense signal (Fig.~\ref{fig:fig4} (a)). The images in Figs.~\ref{fig:fig4} (b) and (c) plot DFXM for  $\beta = 144.74^{\circ}$ and $\beta = 54.74^{\circ}$, showing a compromise between having the wave front perpendicular to the observation plane, which makes the wave appear more narrow and maximizing the projected strain signal given by $\varepsilon \cos(\beta)$, i.e.~the signal that DFXM measures.

The contrast exhibited in the  $\chi$ scans is less pronounced. As already discussed the applicability of geometrical optics for strong-beam conditions is questionable, making this scan geometry better described using the wave-optics formalism, as discussed in the next section.

\subsubsection{Wave optics formalism}
\begin{figure}
\centering
\includegraphics[width = 0.7\textwidth]{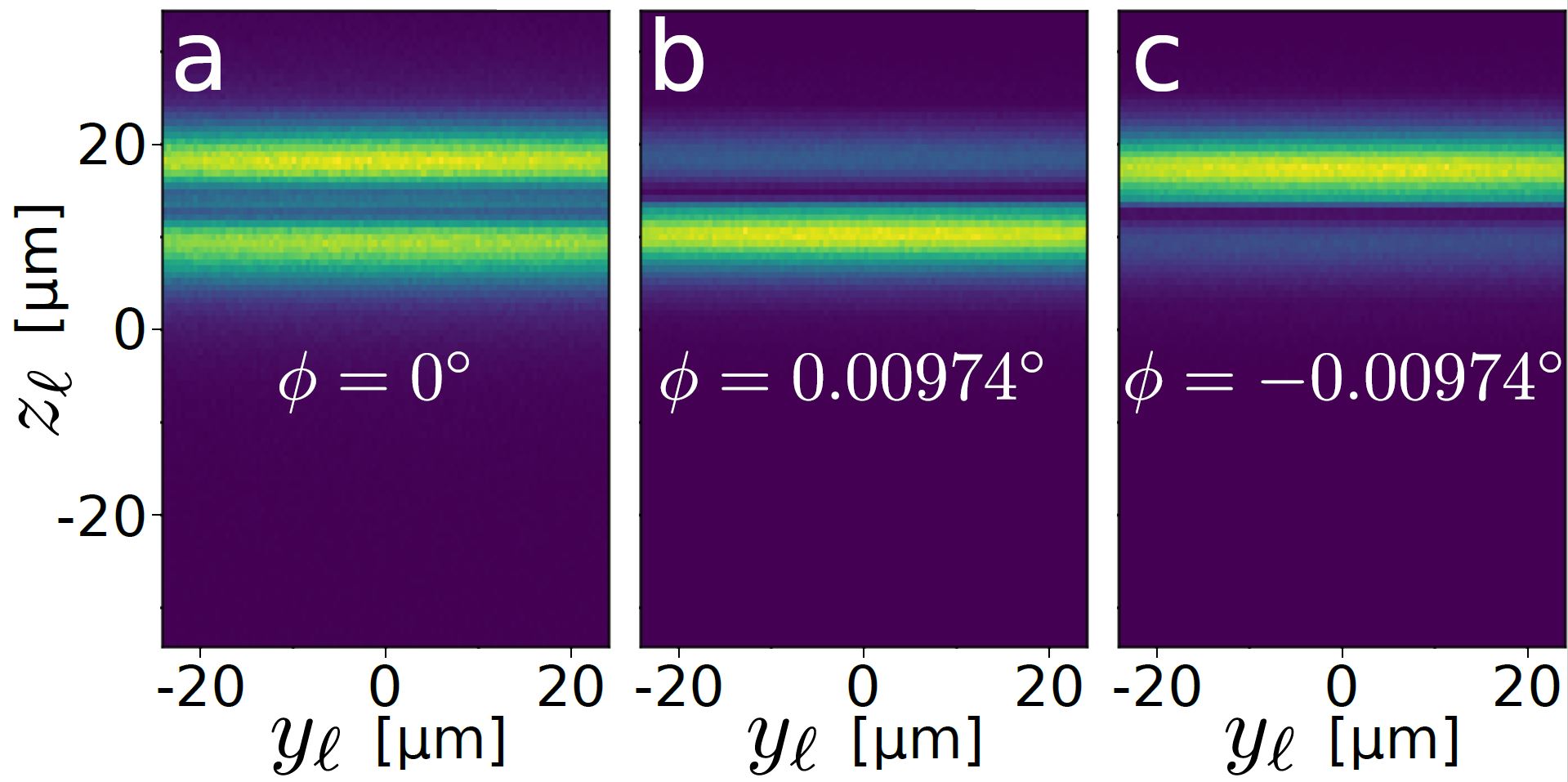}
\caption{Results from wave-optics simulations in the (a) strong-beam, and $\phi$-type weak-beam from (b) rocking in a positive and (c) negative direction. All simulations are done with the sound waves propagating along the [$\bar{1}$10] direction ($\beta$~=~144.74$^\circ$), and using the strain profile shown in Fig.~\ref{fig:fig2} translated to the center of the crystal. The intensity is shown with arbitrary units.
%~\cite{Carlsen2021a}).
}
\label{fig:wave_optics_res}
\end{figure}

Results from the wave-optics simulations are shown in Figure \ref{fig:wave_optics_res}. These simulations are constructed with the strain profile in Fig.~\ref{fig:fig2} (a) propagating along the [$\bar{1}$10] direction ($\beta = 144.74^\circ$; see Figs.~\ref{fig:fig1} and \ref{fig:fig4} (e)). As in the previous example, we plot the strain-wave at a time when it is at the center of the FoV. Figure \ref{fig:wave_optics_res} (a) gives a realistic view of the strong-beam condition, in which we account for dynamical diffraction that causes the strain-wave to appear blurred.

In the strong-beam condition, the bulk shows low intensity as the low-frequency Fourier components (corresponding to the rays at a low divergence angle) of the incoming beam are multiply scattered and spread across the entire FoV. The defect, on the other hand, scatters the high-frequency components of the beam, which are unlikely to scatter again, and thus appears brighter. The center of the strain wave is, however, too far from the diffraction condition to scatter strongly. The FoV of this image is taken from the center of the crystal, where the periodicity of the so-called Pendellösung fringes (from dynamical scattering) is larger than the FoV and appears as a constant background. The intensity of the Pendellösung fringes is much smaller than the intensity generated by the strain wave, but may provide a challenge to imaging if the strain-wave has a smaller amplitude.

In comparison, to get from Fig.~\ref{fig:wave_optics_res} (a) to (b) and (c), the sample is rocked into the $\phi$-type weak-beam condition, causing the strain-wave to appear as a much more clearly defined object in the DFXM images. 
This can be rationalized by considering the contrast mechanisms in the two cases.
When rocking the sample, the crystal is rotated out of the Bragg condition while the strain wave introduces a longitudinal distortion which both elongates and rotates the scattering vector.
This rotation brings the highly-strained center of the strain wave closer to the Bragg condition than the surrounding crystal. Specifically, a positive $\phi$ brings the compressive part of the strain wave closer to the Bragg condition (Fig.~\ref{fig:wave_optics_res} (b)) while a negative  $\phi$ will bring the tensile part of the strain wave closer to the Bragg condition (Fig.~ \ref{fig:wave_optics_res} (c)).
%This gives stronger scattering at the center of the strain-wave.
%Since the observable part of the strain-wave is thinner than the width of the beam, the definition of the strain-wave in Figs.~\ref{fig:wave_optics_res} (b,c) is largely determined by the beam profile.

\section{Discussion and outlook}
The above simulations demonstrate (i) that we can generate strain waves in diamond by using an optical laser to heat a deposited gold film, and (ii) that it should be possible to perform single pulse DFXM imaging of the movement of these strain waves at an XFEL, demonstrating the feasibility of this approach. 

This new approach will also introduce a pump-probe geometry to the DFXM experiment, as opposed to the real-time imaging that has been performed previously \cite{Dresselhaus-Cooper2020, Dresselhaus-Marais2021}. A successful transfer of the DFXM methodology to pump-probe studies at XFELs would imply that the time resolution of diffraction based X-ray imaging can be improved by 9 orders of magnitude, at least for reversible phenomena. 

%When rolled, strong scattering is still observed, but the dynamical part of the intensity is no longer captured by the detector, typically giving an appearance similar to a 'weak-beam'. However, the equivalent shear distortion is now orthogonal to the rotation direction of the scattering vector when distorted by the strain wave, and the contrast when rolling therefore mimics the contrast in Figure \ref{fig:wave_optics_res}a.
%Rolling the sample brings inverted contrast compared to rolling, as the bulk of the sample is again brought to the weak-beam contrast, and in this case the strain does not bring the strained part of the crystal back to the diffraction condition.

%Similar to geometrical optics results. 

%Discuss dynamical effects and effects of coherence. 

In future experiments, the pump-probe measurment scheme can be combined with scanning the goniometer ($\omega$, $\phi$ and $\chi$; see Fig.~\ref{fig:fig1}) to map out the picosecond by picosecond strain-wave-induced dynamics of the strain tensor in 3D with sub-micrometer spatial resolution.

To demonstrate the contrast we have relied on two DFXM forward simulations tools \cite{Poulsen2021, Carlsen2021a}. This work corroborates the validity of these models, as -- although they were constructed independently, using different physical approaches, they arrive at qualitatively similar results. This result provides confidence in the contrast mechanisms proposed, and in our approach to planning experiments for the XFEL capabilities. The comparison also illustrates that dynamical effects do not prohibit the visualisation of a strain wave. During this work it also became clear that the geometrical optics code is well suited for optimization-purposes as the code runs fast - allowing a rapid exploration of the parameter-space. The wave-optics based simulations, while slower, are superior in terms of their ability to explore dynamical diffraction effects. %[Is this code available for others? If you include the code and/or your resolution functions for the methods in the SI, you'll get more interest and citations from people who are inspired to try out this cool new technique.] 

The calibration of the DFXM images simulated in this work to the incident beam flux, i.e. Equation (\ref{eq-intensity-cal}), is approximate at this time, as it neglects factors such as beam polarization, attenuation of beams along optical axes in the X-ray optics, vignetting and quantum efficiency of the detector that can also contribute to the DFXM signal. Perhaps more importantly, positional jitter of the incident beam may reduce intensities. Such positional jitter can be challenging in multiple ways. (i) If a monochromator and/or upstream cleanup slit are introduced, the positional jitter manifests in intensity jitter at the sample because the relevant apertures clip the beam differently as the beam positionally shifts. (ii) The positional jitter can also change the pointing of the beam, changing the incident wavevectors and corresponding diffraction condition, making it hard to selectively focus on one orientation or lattice-spacing. (iii) Sufficient positional jitter can also make CRLs prohibitively difficult to align \cite{Breckling2021}. 

Having a stable photon energy is also important. If the photon energy or bandwidth jitter, the resulting DFXM beam will sample a different lattice-spacing on each shot. Since the self-amplified spontaneous emission beam amplifies regions of the XFEL bandwidth randomly, it shifts the photon energies that determine contrast in DFXM images non-uniformly. This introduces overlaid components of the lattice-spacings in a way that is hard to deconvolve when interpreting the results.

On the other hand, the expression for the ratio of intensity being dynamically diffracted may be conservative as the mosaic spread  may be smaller locally within the crystal. Moreover, these effects can be readily simulated using the approaches outlined above. Given the count rates of the order of 100 in our simulated images, we conclude that single pulse visualisation of the strain wave with an LCLS type set up is feasible. 

In the XCS setup used for these simulations, the long focal length of the condenser at the beamline (3 m) puts a limit on the beam width. Experiments could overcome this by placing the condenser closer to the sample, facilitating beam heights of a few hundred nanometers. We note, however, that this effect is not as simple as just moving the focus, as it introduces light that focuses over a significantly shorter distance, introducing a large change in the divergence of the incident beam and the depth-of-focus that describes the observation plane through the depth of the crystal. Given the enhanced scope of that modification, it is beyond the scope of this initial study. %The results presented suggests that in terms of signal-to-noise  an in plane spatial resolution of order 200 nm is  within reach. 

\section{Conclusion }

In this work, simulations based on both geometrical optics and wave optics were used to demonstrate the feasibility of DFXM imaging of laser-generated strain waves in diamond single crystals using an XFEL-source. Besides showing that these optics-formalisms give consistent results, and discussing their strengths and weaknesses, these simulations paint an optimistic picture for upcoming experiments at the XCS beamline at LCLS. These experiments, if successful, would constitute a major advancement for DFXM. This would, in turn, open the door for the study of a plethora of ultrafast phenomena, such as, e.g., interactions between strain waves and defects (e.g. dislocations, twin walls and grain boundaries), rapid material failure and diffusionless transformations. 

\iffalse
\appendix
\section{Appendix title}

Text text text text text text text text text text text text text text
text text text text text text text.

\subsection{Title}

Text text text text text text text text text text text text text text
text text text text text text text.

\subsubsection{Title}

Text text text text text text text text text text text text text text
text text text text text text text.

\fi
     %-------------------------------------------------------------------------
     % The back matter of the paper - acknowledgements and references
     %-------------------------------------------------------------------------

     % Acknowledgements come after the appendices

\subsection*{Acknowledgements}
 We are grateful to  Bernard Kozioziemski, Eric Folsom, Matthew H. Seaberg , John H. Eggert and  Tim van Driel for scientific discussions and technical specifications for the set-up at LCLS.
We are very grateful to our main sponsor, the Villum Foundation, which has enabled us to carry out this work. 
In addition, HFP acknowledges an ERC Advanced Grant, nr 885022 and support from the ESS lighthouse on hard materials in 3D, SOLID, funded by the Danish Agency for Science and Higher Education, grant number 8144-00002B. 

Partial contributions from LEDM were performed under the auspices of the U.S. Department of Energy by Lawrence Livermore National Laboratory under Contract DE-AC52-07NA27344, and the Lawrence Fellowship.

We are also very grateful to D. Schick for useful discussions, and for developing and maintaining the code (\texttt{udkm1Dsim}) used for strain wave simulations.  
     % References are at the end of the document, between \begin{references}
     % and \end{references} tags. Each reference is in a \reference entry.

     %-------------------------------------------------------------------------
     % TABLES AND FIGURES SHOULD BE INSERTED AFTER THE MAIN BODY OF THE TEXT
     %-------------------------------------------------------------------------

     % Simple tables should use the tabular environment according to this
     % model

\bibliography{refs}{}
\bibliographystyle{unsrt}

\end{document}